# When the quarter jumps into a cup (and when it does not)

Mateo Dutra[1], Álvaro Suárez[2], Martín Monteiro[3], Arturo C. Marti[1]

[1]Instituto de Física, Universidad de la República, Uruguay

[2]Consejo de Formación en Educación, Montevideo, Uruguay

[3]Universidad ORT Uruguay, Uruguay

While Bernoulli's equation is one of the most frequently mentioned topics in Physics literature and other means of dissemination, it is also one of the least understood. Oddly enough, in the wonderful book *Turning the world inside out*[1], Robert Ehrlich proposes a demonstration that consists of blowing a quarter dollar coin into a cup, incorrectly explained using Bernoulli's equation. In the present work, we have adapted the demonstration to show situations in which the coin jumps into the cup and others in which it does not, proving that the explanation based on Bernoulli's is flawed. Our demonstration is useful to tackle the common misconception, stemming from the incorrect use of Bernoulli's equation, that higher velocity *invariably* means lower pressure.

**When the quarter jumps into the cup.** The experiment proposed by Ehrlich, illustrated in Figure 1, consists of blowing across the surface of a coin so that it jumps into a cup. In his explanation, the author states that the pressure of the air when blowing must decrease as a consequence of Bernoulli's equation, which makes the coin jump. Matching the lift force to the weight of the coin, the critical velocity of the blow (assumed to be uniform) for the coin to jump turns out to be

$$v_{cr} = \sqrt{\frac{2mg}{\rho A}},$$

where $m$ and $A$ stand for the mass and surface of the coin, respectively, and $\rho$ is the air density. For the common varieties of coins in circulation within the USA, this value is between 12 and 15 m/s, of the order of the typical velocity of a blow. Now, what is wrong with that explanation? Bernoulli's equation has several restrictions, one of them being that the application points must be in a single streamline. Several published works propose laboratory experiments or discuss the consequences of "abusing" Bernoulli's equation[2,3,4].

**Adapting the demonstration.** We present an adaptation of the coin experiment that makes it clear that the lift force that raises the coin is attributable to the curvature of the streamlines rather than to the assumption that "higher velocity means lower pressure". The experiment shown in Figure 2 consists of placing the coin in a gap in the table's surface so that the top of the coin is level with the surface (*b*), instead of simply placing it on the table (*a*). A piece of cardboard of the same thickness as the coin can be used as the surface, making a circular hole in it a little larger than the coin's size. The experimental results showed us that it is relatively easy to make the coin jump with a normal blow in setup *a*, but we were not able to do so in setup *b* despite our most energetic blowing efforts. Even when resorting to a blower such as those used for Physics laboratory air tracks, we arrived at the same results.

In conclusion, in setup *a*, the lift force stems from the curvature of the streamlines that, according to Newton's third law, produces a lift force on the coin. The reasoning based on Bernoulli's equation is flawed because, while it can explain the jump in *a*, it contradicts the experimental results in *b*. Therefore, with this low-cost experiment we show that higher velocity in fluids does not always mean lower pressure. This demonstration is ideal for use in a PODS (prediction, observation, discussion and synthesis) mode in introductory fluid dynamics courses.

**FIGURES**

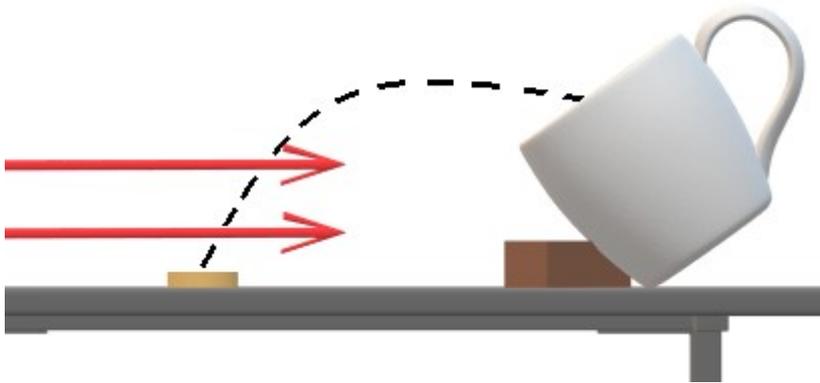

**Figure 1.** Ehrlich experiment. "By blowing across the surface of a coin, you can make it jump into a cup, thus illustrating Bernoulli's principle and allowing you to calculate the minimum velocity of the blow."

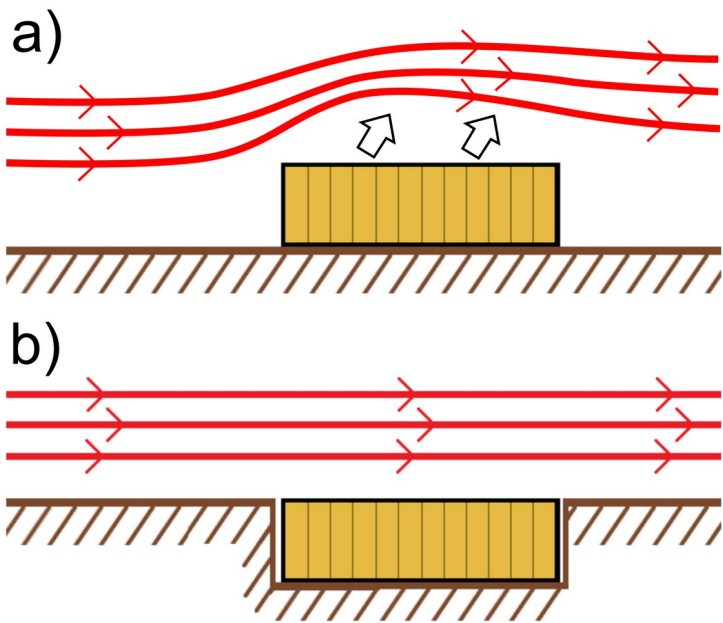

**Figure 2.** When blowing on a coin placed on top of the table (a), the coin jumps easily, whereas if the coin's surface is level with the table (b), it does not lift, no matter how hard you blow. In the first case, the streamlines are deflected and a lift force on the coin is generated pursuant to Newton's third law. In the second case, the streamlines are not deflected and, therefore, no lift force is generated. A reasoning based on Bernoulli could be applied in both cases, but in the second case it would contradict the experimental results.